\documentclass[%
 reprint,
 showkeys,
 amsmath,amssymb,
 aps,
 prb,
 superscriptaddress,
 nobibnotes
]{revtex4-1}

\usepackage{graphicx}
\usepackage{dcolumn}
\usepackage{bm}

\makeatletter
\AtBeginDocument{\@ifpackageloaded{natbib}{\ifNAT@numbers\if@filesw\immediate\write\@auxout{\string\global\string\NAT@numberstrue}\fi\fi}{}}
\makeatother
\usepackage{amsmath}

\begin{document}


\title{Zigzag Chain Structure Transition and Orbital Fluctuations in Ni-based Superconductors}

\author{Youichi Yamakawa}
 \email[]{yamakawa@s.phys.nagoya-u.ac.jp}
\affiliation{Department of Physics, Nagoya University, Furo-cho, Nagoya 464-8602, Japan }
\author{Seiichiro Onari}
\affiliation{Department of Applied Physics, Nagoya University, Furo-cho, Nagoya 464-8603, Japan}
\author{Hiroshi Kontani}
\affiliation{Department of Physics, Nagoya University, Furo-cho, Nagoya 464-8602, Japan }



\date{\today}

\begin{abstract}
 We investigate the electronic state and structure transition of BaNi$_2$As$_2$, 
  which shows a similar superconducting phase diagram as Fe-based superconductors. 
 We construct the ten-orbital tight-binding model for BaNi$_2$As$_2$ by using the maximally localized Wannier function method.
 The Coulomb and quadrupole-quadrupole interactions are treated within the random-phase approximation.
We obtain the strong developments of charge quadrupole susceptibilities
driven by the in-plane and out-of-plane oscillations of Ni ions.
The largest susceptibility is either 
$O_{X^2-Y^2}$-quadrupole susceptibility at ${\bm q} = (\pi, 0, \pi)$
or $O_{XZ(YZ)}$-quadrupole susceptibility at ${\bm q} = (\pi, \pi, \pi)$,
depending on the level splitting between $d_{\rm X^2-Y^2}$ and $d_{\rm XZ(YZ)}$.
 These antiferro-quadrupole fluctuations would then be the origin of the strong coupling superconductivity in Ni-based superconductors. 
 Also, we propose that the antiferro-quadrupole $O_{X^2-Y^2}$ order with ${\bm q} = (\pi, 0, \pi)$ is the origin of the zigzag chain structure reported in experiments.
 We identify similarities and differences between Ni- and Fe-based superconductors. 
\end{abstract}

\maketitle

\section{Introduction}
 Since the discovery of superconductivity in the Fe-based pnictide by Kamihara {\it et al}., \cite{Kamihara}
  great efforts have been devoted to the discovery of other transition metal based pnictides. 
 Recently, BaNi$_2$(As$_{1-x}$P$_x$)$_2$ attracts increasing attention due to its unique phase diagram, 
  in which strong coupling superconductivity is realized next to the structure transition. \cite{Ronning, Kurita, Subedi, Sefat, Kudo}
 No evidence of spin-density-wave (SDW) order has been observed. 
 Above the structure transition temperature $T_{\rm s} \sim 130$~K,
  it has the same tetragonal structure as BaFe$_2$As$_2$, which is a typical parent compound of Fe-based superconductors.
 At $T_{\rm s}$, BaNi$_2$(As$_{1-x}$P$_x$)$_2$ exhibits a first-order triclinic structure transition, \cite{Ronning, Sefat}
  whereas Fe-based compounds exhibits a second-order orthorhombic transition. \cite{Rotter, Krellner}
 Below $T_{\rm s}$, the Ni atoms form zigzag chains with shorter Ni-Ni distances ($\sim$~2.8~\AA) and the chains are separated by significantly longer Ni-Ni distances ($\sim$~3.1~\AA). \cite{Sefat}
 The superconductivity emerges at $T_{\rm c} \sim 0.7$~K in this triclinic phase, 
  which is considered to be of the conventional weak-coupling BCS type. \cite{Ronning, Kurita, Subedi, Kudo}
 From a previous study of density functional theory (DFT) calculations for BaNi$_2$As$_2$, \cite{Subedi}
  it was found that a strong electron-phonon interaction ($\lambda_{\rm ep} = 0.76$) is induced by the low-energy ($\sim \negthickspace50$~K) Ni and As oscillating modes.
 Kudo {\it et al.} recently reported the suppression of the structural transition in BaNi$_2$(As$_{1-x}$P$_x$)$_2$ and the emergence of a new strong-coupling superconducting state ($\Delta C/\gamma T_{\rm c} \sim 2$) with $T_{\rm c} \ge 3.3$~K for $x > 0.07$. \cite{Kudo}

 The undoped Fe-based superconductors also exhibit structural transition from tetragonal to orthorhombic at $T_{\rm s}$ and SDW transitions at $T_{\rm N}$. \cite{Cruz, Klauss, Johnston, Rotter, Krellner, Huang}
 In many compounds, $T_{\rm s}$ is higher than $T_{\rm N}$, and $T_{\rm s}$ is expected to be realized by orbital polarization. \cite{Lv, Lee, Onari_VC, Kontani_C66}
 In both electron- and hole-doped cases, superconductivity appears near the structure, orbital, and SDW orders. \cite{Kamihara, Johnston, Cruz, Rotter_SC, Sefat_SC}
 It is, therefore, natural to consider the possibility that the orbital and spin fluctuations provide the pairing mechanism,
  and two types of superconducting states are proposed:
  $s_{\pm}$ superconductivity mediated by the spin fluctuation \cite{Kuroki, Graser, Chubukov} and $s_{++}$ superconductivity mediated by the orbital fluctuation. \cite{Kontani_SC, Saito, Onari_FLEX, Kontani_001}
 In many compounds, the superconductivity is very robust against impurities, \cite{Sato, Nakajima, Li} indicating that the $s_{++}$-wave state is realized in these compounds. \cite{Onari_Impurity} 

 In Fe-based superconductors, a large softening of the elastic constant $C_{66}$ has been reported above $T_{\rm s}$. \cite{Fernandes, Yoshizawa, Goto}
 Also, spontaneous in-plane anisotropy of the electronic states, 
  so called the nematic state, 
  was reported by the magnetic torque \cite{Kasahara} and resistivity measurements. \cite{Chu, Nakajima_Nematic} 
 These facts indicate the existence of
  large orbital fluctuations, and the relationship between the orbital fluctuations and superconductivity has attracted much attention. 
 However, the softening and the relation $T_{\rm s} > T_{\rm N}$ can not be explained within the random phase approximation (RPA) for multi-orbital Hubbard model. \cite{Kuroki}
 Recently, we have improved the RPA by including the vertex corrections (VCs), 
  and found that the strong ferro- and antiferro-orbital fluctuations are caused by the strong spin-orbital coupling described by the VC. \cite{Kontani_C66, Onari_VC}
 The obtained orbital fluctuations well explain both the structure transition, including the large $C_{66}$ softening, and $s_{++}$-wave superconducting state. 
 
 For BaNi$_2$(As$_{1-x}$P$_x$)$_2$, a large phonon softening toward the structure transition has also been reported, \cite{Kudo}
  although the structure transition is of first order. 
 Therefore, strong orbital fluctuations are expected to exist in BaNi$_2$(As$_{1-x}$P$_x$)$_2$ as in Fe-based superconductors.
 It is noteworthy that Ir$_{1-x}$Pt$_x$Te$_2$ \cite{Pyon} and CaC$_6$ \cite{Kim, Gauzzi} also exhibit superconducting transitions and a lowering of phonon frequencies next to the structure transitions.
 Therefore, it is important to understand the relationship among  structure transition, orbital fluctuations, and superconductivity.
 It is also significant to identify similarities and differences between Ni- and Fe-based superconductors. 

 Herein, we investigate 
the dynamical spin and orbital susceptibilities and discuss the phase transitions
 of BaNi$_2$As$_2$ based on RPA.
 We construct the ten-orbital tight-binding model for BaNi$_2$As$_2$ by using the maximally localized Wannier function method. 
 The obtained critical value of the Coulomb interaction for the SDW state, $U_{\rm c}$, is very large, 
  which is consistent with the absence of the SDW state. 
 However, it is found that the quadrupole interaction resulting from Ni-ion oscillations gives rise to the antiferro-quadrupole (AFQ) order, 
  which presents a natural explanation for the zigzag chain structure in the triclinic phase. 
 We also discuss the orbital-fluctuation-driven superconductivity in Ni-based superconductors.

\section{Possible quadrupole order}
\begin{figure}[tb]
        \includegraphics[width=8cm]{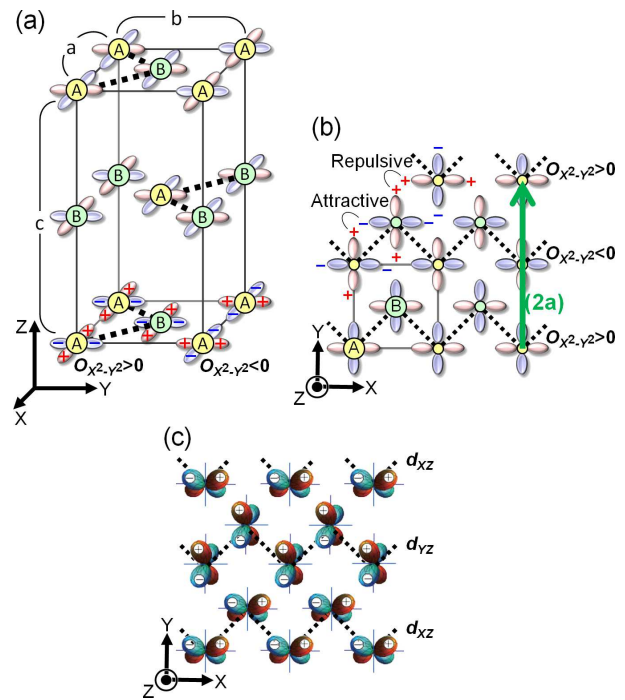}
        \caption{\label{fig1}
                (Color online)
                (a) Crystal structure of BaNi$_2$As$_2$ with AF $O_{X^2-Y^2}$ order with momentum $\bm{q} = (0, \pi, \pi)$.
                 The dotted zigzag lines correspond to the shorter bond.
                 A and B represent the Ni-A and Ni-B ions in a primitive unit cell.
                 The box with a solid line corresponds to a conventional unit cell.
                (b) Two-dimensional picture of the AFQ.
                 The square with a solid line corresponds to a primitive unit cell.
                 The arrow $(0, 2a)$ represents the translation vector corresponding to $\bm{q} = (0, \pi)$.
                (c) The antiferro orbital order, which is equivalent to the AFQ order in (b).
        }
\end{figure}

 Before performing numerical calculations, we discuss the possible quadrupole order that represents the experimental zigzag chain structure. 
 Figure~\ref{fig1}(a) and \ref{fig1}(b) present schematic pictures of the $O_{X^2-Y^2}$ order with momentum $\bm{q} = (0, \pi, \pi)$
 that corresponds to the zigzag chain structure. 
 Here, $a$, $b$, and $c$ are lattice constants, and there are two Ni ions, A and B, per unit cell.
 In this study, we set the $X$ and $Y$ axes  along  the $a$ and $b$ axes (so that the Ni-As direction is in the $a\text{-}b$ plane) and the $Z$ axis is then parallel to the $c$ axis, as shown in Fig.~\ref{fig1}(a).
 The symbols $+$ and $-$ around the Ni ions represent the sign of the charge distribution of the $O_{X^2-Y^2}$ quadrupole.
 $O_{X^2-Y^2}$ is an Ising-type order parameter.
 
 As shown in Fig.~\ref{fig1}(b), the values of $O_{X^2-Y^2}$ on Ni ions have the same sign on each zigzag chain (as shown by dotted lines).
 In this case, the bond length on the zigzag chain would become shorter owing to the electrostatic potential.
 However, Ni ions on different zigzag chains would feel a repulsive force and form a longer bond.
 Experimentally, longer bond (shorter bond) is about 3.1~\AA{} (2.8~\AA{}). \cite{Sefat} 
 Such a large difference between longer and shorter bonds might indicates the importance of electron-phonon interaction. 
 Since $O_{X^2-Y^2} \propto n_{d_{\rm XZ}} - n_{d_{\rm YZ}}$, 
  the quadrupole order $O_{X^2-Y^2} > 0$ corresponds to the orbital polarization $n_{d_{\rm XZ}} > n_{d_{\rm YZ}}$. \cite{Kontani_C66}
 Thus, the AF $O_{X^2-Y^2}$ order is equivalent to the antiferro orbital order shown in Fig.~\ref{fig1}(c),
 in which the mainly occupied $d$-orbital wave functions are shown.
 
 In this paper, we investigate the development of quadrupole fluctuations due to
  quadrupole interaction, which is induced by Ni-ion oscillations.
 We obtain the divergent development of $O_{X^2-Y^2}$ susceptibility at $\bm{q} = (\pi, 0, \pi)$ (and $\bm{q} = (0, \pi, \pi)$), 
  which is consistent with the zigzag structure shown in Fig.~1(a)-(c).

\section{First principle study and model Hamiltonian}
\begin{figure}[tb]
        \includegraphics[width=8cm]{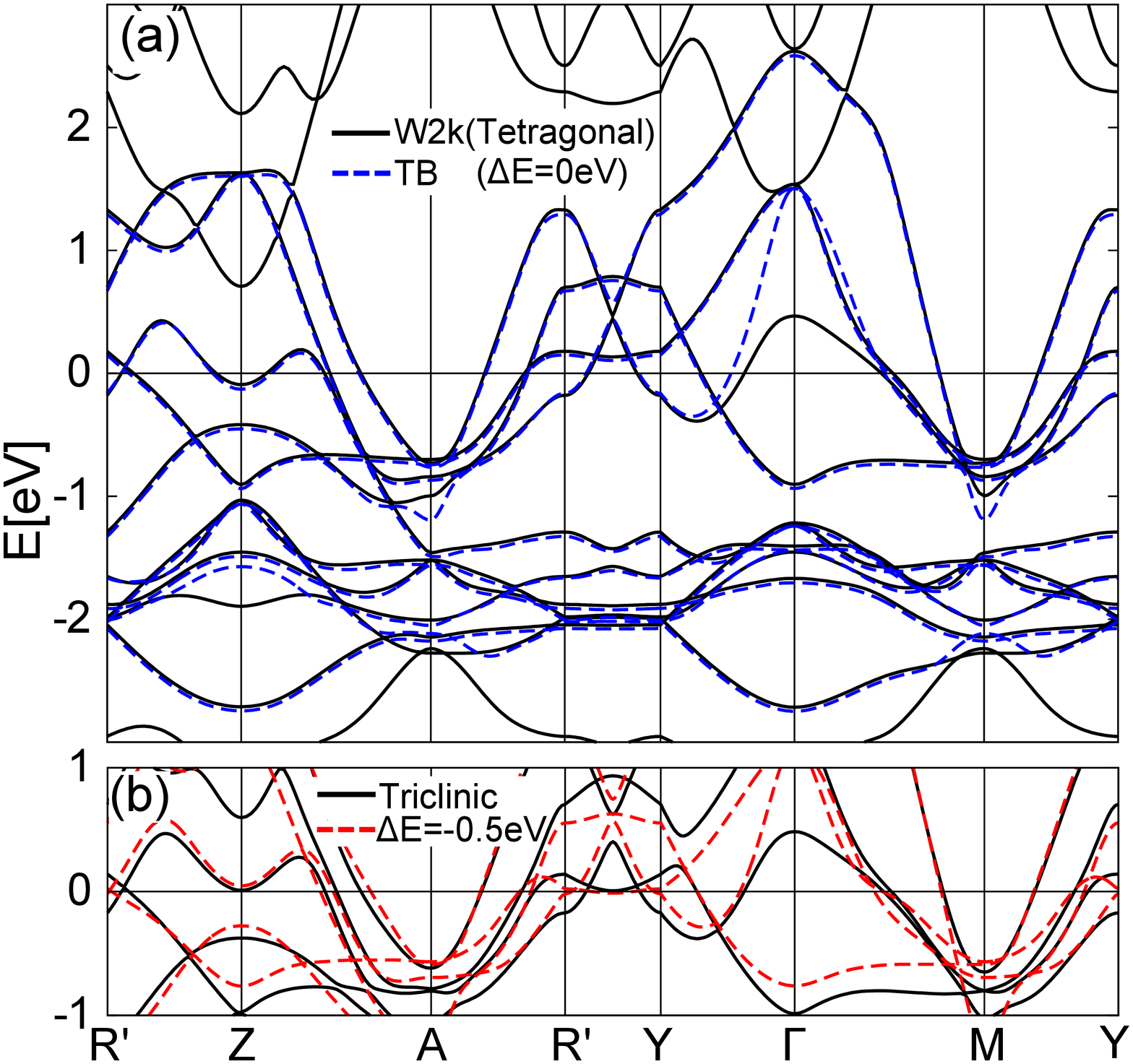}
        \includegraphics[width=6cm]{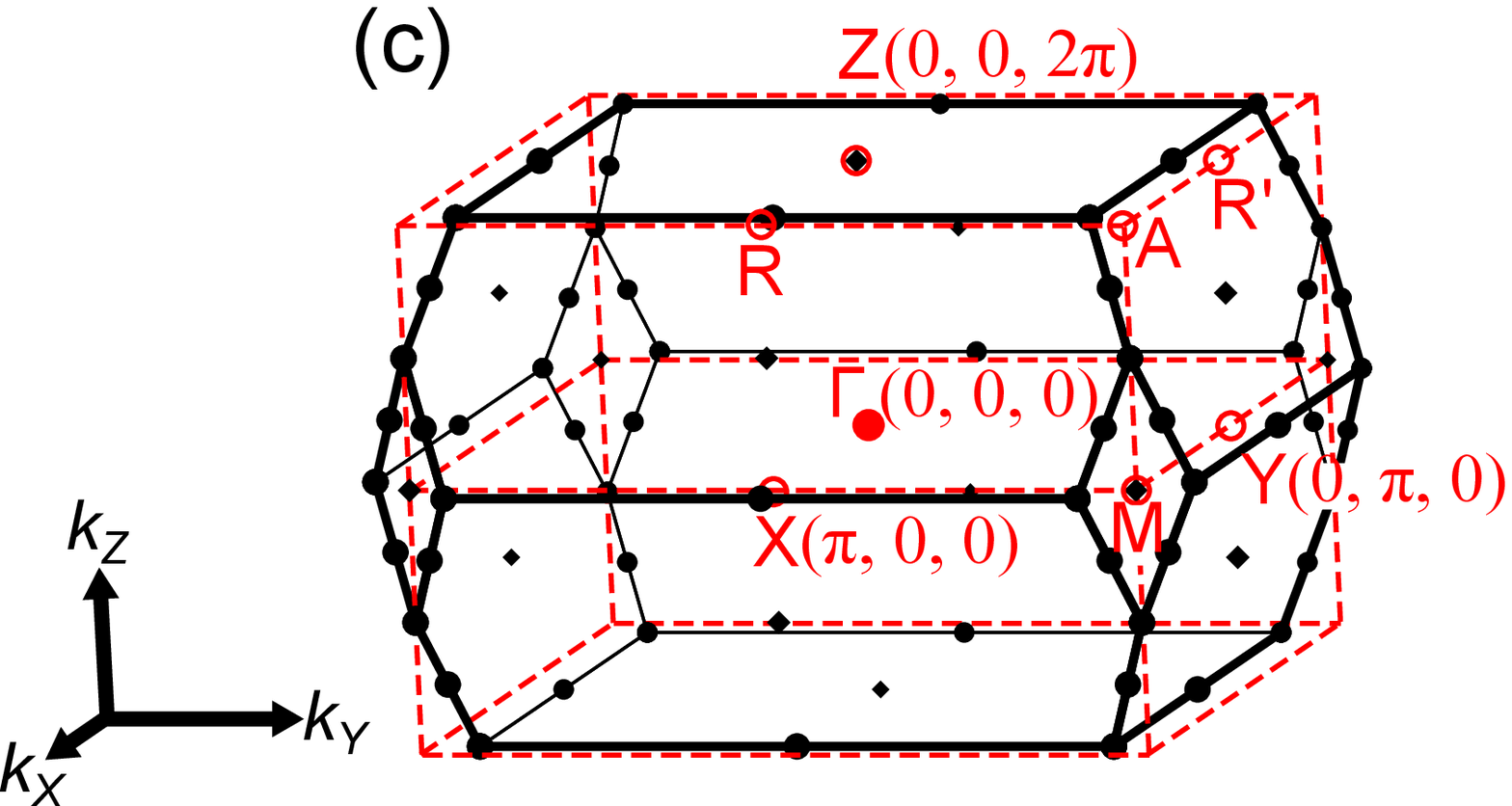}
        \includegraphics[width=8cm]{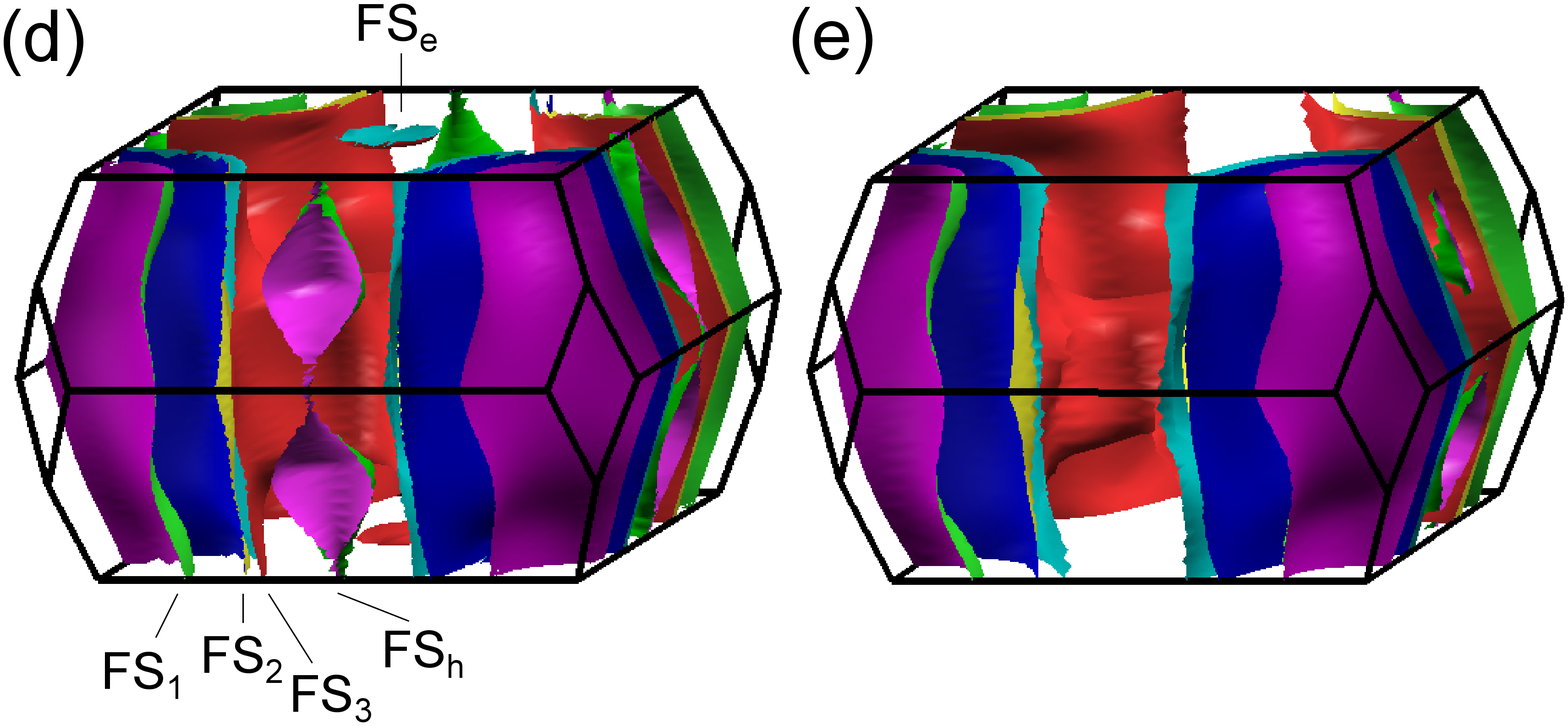}
        \caption{\label{fig2}
                (Color online)
                (a) and (b) Band structure of BaNi$_2$As$_2$.
                 In (a), the solid and dashed lines correspond to the DFT calculation for the tetragonal phase and the tight-binding model for $\Delta E = 0$, respectively. In
                (b), the solid lines correspond to the DFT calculation for the triclinic phase,
                  and dashed lines are given by the tight-binding model for $\Delta E = - 0.5$~eV.
                (c) Brillouin zone.
                (d) Fermi surfaces in the tetragonal phase given using DFT.
                (e) Fermi surfaces in the triclinic phase.
        }
\end{figure}

 First, we perform a DFT calculation for BaNi$_2$As$_2$ with the generalized gradient approximation by using the \textsc{wien2k} package. \cite{wien2k}
 In our calculation, we used experimental lattice parameters. \cite{Sefat}
 Figure~\ref{fig2}(a) and \ref{fig2}(b) show the band dispersions obtained from the DFT calculation (solid lines) on the tetragonal and the triclinic phases, respectively.
 The Brillouin zone is shown in Fig.~\ref{fig2}(c).
 The band structure of BaNi$_2$As$_2$ in the tetragonal phase is qualitatively similar to that for BaFe$_2$As$_2$,
  although the Fermi level is shifted upward since Ni$^{2+}$ contains two more valence electrons than Fe$^{2+}$.
 The obtained result is consistent with those of previous studies of DFT calculations. \cite{Subedi, Shein, Chen, Zhou}
 As a result of this upward shift, BaNi$_2$As$_2$ exhibits very different electronic properties from those of BaFe$_2$As$_2$.
 The Fermi surfaces in tetragonal and triclinic phases are shown in Fig.~\ref{fig2}(d) and \ref{fig2}(e), respectively.
 As shown in Fig.~\ref{fig2}(d), there are three large Fermi surfaces FS$_1$--FS$_3$,
  a small dishlike electron pocket FS$_{\rm e}$ around $Z$ point $(0, 0, 2\pi),$ and a hole pocket FS$_{\rm h}$ around $(\pi, 0, \pi)$.
 FS$_1$ and FS$_2$ have simple cylindrical forms,
  whereas FS$_3$ has a very complex structure.
 The shape of these Fermi surfaces is similar in both tetragonal and triclinic phases.
 In contrast, FS$_{\rm e}$ disappears in the triclinic phase, as shown in Fig.~\ref{fig2}(e).
 Further, FS$_{\rm h}$ disappears around ($\pm \pi, 0, \pm \pi$) owing to the energy gap opening up as the symmetry is lowered, and it becomes small at around ($0, \pm \pi, \pm \pi$).
 We note that the $X$ and $Y$ axes are not equivalent in the triclinic phase.
 
 Next, we introduce the ten-orbital tight-binding model for BaNi$_2$As$_2$ with the tetragonal structure.
 The tight-binding Hamiltonian is given by
\begin{eqnarray}
        \hat{H}^{0}
        &=&
                \sum_{i}
                \sum_{\alpha, l, \sigma}
                \epsilon_{l}
                c_{\alpha, l, \sigma}^{i \, \dagger}
                c^{i}_{\alpha, l, \sigma}
        \nonumber \\
        &&+     \sum_{i, j}
                \sum_{\alpha, \beta, l, m, \sigma}
                t^{i, j}_{\alpha, \beta, l. m}
                c_{\alpha, l, \sigma}^{i \, \dagger}
                c^{j}_{\beta, m, \sigma}
        ,
\end{eqnarray}
where $i$ and $j$ denote the unit cell,
  $\alpha$ and $\beta$ represent Ni ions A and B,
  $l$ and $m$ denote the $d$ orbital ($d_{3Z^2 - R^2}$, $d_{XZ}$, $d_{YZ}$, $d_{X^2-Y^2}$, or $d_{XY}$),
  and $\sigma = \pm 1$ is the spin index.
$c_{\alpha, l, \sigma}^{i \dagger}$ ($c^{j}_{\beta, m, \sigma}$) is a creation (annihilation) operator of the Ni $3d$ electron.
 The on-site energies $\epsilon_{l}$ and hopping integrals $t^{i, j}_{\alpha, \beta, l, m}$ are obtained from maximally localized Wannier functions using the \textsc{wannier90} code \cite{wannier90} and the \textsc{wien2wannier} interface. \cite{wien2wannier}

\begin{figure}[tb]
        \includegraphics[width=8cm]{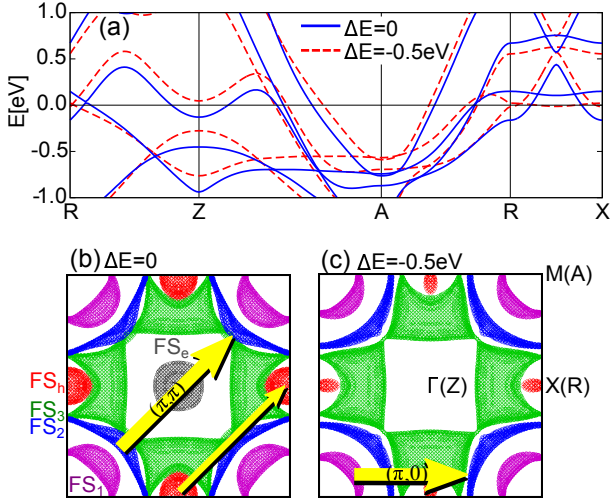}
        \caption{\label{fig3}
                (Color online)
                (a) The band structure of the tight-binding model for $\Delta E = 0$ and $\Delta E = - 0.5$~eV.
                (b) and (c) The Fermi surfaces of the tight-binding model for $\Delta E = 0$ and $\Delta E = - 0.5$~eV projected onto the $k_X$-$k_Y$ plane.
                 The arrows $(\pi, \pi)$ and $(\pi, 0)$ are the nesting vectors that correspond to the peak of the quadrupole susceptibilities shown in Fig.~\ref{fig5}.
        }
\end{figure}

 We make comparison of the band structures given by the DFT calculation and the present tight-binding model. 
 In Fig.~\ref{fig2}(a), dashed lines represent the band dispersions of the obtained tight-binding model.
 After the triclinic structure transition, band dispersions show (i) energy level splitting with symmetry lowering
 and (ii) an orbital-dependent energy level shift without symmetry lowering, reflecting the first-order transition.
 We found that the
 latter change can be roughly reproduced by energy level lowering of $d_{XZ}$ and $d_{YZ}$ orbitals, $\Delta E \equiv \Delta E_{d_{XZ}} = \Delta E_{d_{YZ}} = -0.5$~eV.
 The obtained tight-binding dispersions for $\Delta E = -0.5$~eV are shown in Fig.~\ref{fig2}(b) by dashed lines.
 (Note that the tetragonal symmetry is not violated by $\Delta E$.)
 In both band structures in Fig.~\ref{fig2}(b), 
 FS$_{\rm h}$ and FS$_{\rm e}$ in the tetragonal band structure become very small. 
 Therefore, the dominant inter-orbital nesting vector relevant for the orbital fluctuations is expected to be given by the arrow in Fig.~\ref{fig3}(c).

 For comparison, we show the band structure for $\Delta E = 0$ and $\Delta E = - 0.5$~eV in Fig.~\ref{fig3}(a).
 We also show the Fermi surfaces (projected onto the $k_X$-$k_Y$ plane) for $\Delta E = 0$ and $\Delta E = - 0.5$~eV in Fig.~\ref{fig3}(b) and \ref{fig3}(c), respectively.
 In Fig.~\ref{fig3}(c), FS$_{\rm e}$ disappears and FS$_{\rm h}$ shrinks, consistently with the DFT band dispersion for the triclinic phase.
 Hereafter, we calculate both cases $\Delta E = 0$ and $\Delta E = - 0.5$~eV, and we discuss the spin and quadrupole fluctuations.
 
\section{Numerical results by RPA}
\subsection{Formalism}
 Here, we present the numerical results obtained by RPA. 
 The noninteracting susceptibility for zero frequency is given by
\begin{eqnarray}
        \chi^{0 \, \alpha, \beta}_{l,l',m,m'} ( \bm{q} )
        &=&
        -       \frac{1}{N}
                \sum_{b, b', \bm{k}}
                \frac{
                                f ( \epsilon_{b } (\bm{k} + \bm{q}) )
                          - f ( \epsilon_{b'} (\bm{k}) )
                         }
                         {      \epsilon_{b } (\bm{k} + \bm{q})
                          -     \epsilon_{b'} (\bm{k}) }
\nonumber \\
        && \times
                u_{\alpha, l , b } (\bm{k} + \bm{q}) u_{\beta,  m , b }^{*} (\bm{k} + \bm{q})
\nonumber \\
        && \times
                u_{\beta,  m', b'} (\bm{k})          u_{\alpha, l', b'}^{*} (\bm{k})
        ,
\end{eqnarray}
where $\epsilon_{b} (\bm{k})$ and $f(\epsilon_{b} (\bm{k}))$ are the dispersion and the Fermi distribution function of a quasiparticle for band $b$ with momentum $\bm{k}$.
 The function $u_{\alpha, l, b} (\bm{k}) = \left< \alpha, l | b, \bm{k} \right>$ connects the orbital and the band spaces, given by the procedure for diagonalization of $H^{0}$.

 Now, we analyze the effects of the Coulomb and quadrupole-quadrupole interactions using the RPA.
 The on-site Coulomb interaction is composed of the intra-orbital term $U$, the inter-orbital one $U'$, Hund's coupling $J,$ and pair hopping $J'$.
 We set these parameters as $U' = U - 2J$ and $J' = J$ to satisfy the rotational invariance of the electron-electron interaction term.
 The bare four-point vertex for the spin channel is given by
\begin{eqnarray}
        \Gamma^{\rm s \, \alpha, \alpha}_{l, l', m, m'}
        =
        \left\{
                \begin{array}{lllll}
                        U , & \qquad l & = l' &=   m  & = m', \\
                        U', & \qquad l & = m  &\ne l' & = m', \\
                        J , & \qquad l & = l' &\ne m  & = m', \\
                        J', & \qquad l & = m' &\ne l' & = m.  \\
                \end{array}
        \right.
\end{eqnarray}
 Further, the bare four-point vertex for the charge channel is given by
\begin{eqnarray}
        \hat{\Gamma}^{\rm c}
        =
        - \hat{C} - 2 ( \hat{V}_{\rm quad} )
\end{eqnarray}
  with
\begin{eqnarray}
        C^{\alpha, \alpha}_{l, l', m, m'}
        =
        \left\{
                \begin{array}{lllll}
                        U     , & \quad l & = l' &=   m  & = m', \\
                        -U'+2J, & \quad l & = m  &\ne l' & = m', \\
                        2U'- J, & \quad l & = l' &\ne m  & = m', \\
                        J'    , & \quad l & = m' &\ne l' & = m,  \\
                \end{array}
        \right.
\end{eqnarray}
  and
\begin{eqnarray}\label{eq6}
        ( V_{\rm quad} )^{\alpha, \alpha}_{l, l', m, m'}
        &=&
                - g
                \left(
                        o_{XZ}^{l, l'} o_{XZ}^{m, m'}
                +       o_{YZ}^{l, l'} o_{YZ}^{m, m'}
                \right)
        \nonumber \\
        &&
                - g'
                o_{X^2-Y^2}^{l, l'} o_{X^2-Y^2}^{m, m'}
        ,
\end{eqnarray}
  where $g$ and $g'$ are the charge quadrupole-quadrupole coupling constants.
 In Eq.~(\ref{eq6}) the quadrupole interactions for channels $XZ (YZ)$ and $X^2-Y^2$ are mediated by in-plane and out-of-plane oscillations of Ni ions, respectively. \cite{Kontani_SC, Saito, Kontani_C66, Onari_FLEX}
 The charge quadrupole operator is defined as $\hat{O}^{i, \alpha}_{\Gamma} = \sum_{l, m, \sigma} o^{l, m}_{\Gamma} c^{i \, \dagger}_{\alpha, l, \sigma} c^{i}_{\alpha, m, \sigma}$, 
  and the coefficient $o_{\Gamma}$ is defined as $o_{XZ}^{l, m} = 7 \left< l \left| \frac{X}{R} \frac{Z}{R} \right| m \right>$ for $\Gamma = XZ$.

 In the RPA framework, the spin (charge) susceptibility is obtained by using $\hat{\Gamma}^{\rm s(c)}$ as \cite{Takimoto}
\begin{eqnarray}
        \hat{\chi}^{\rm s (c)} ( \bm{q} )
        =
                \frac{ \hat{\chi}^{0} ( \bm{q} ) }
                         { 1 - \hat{\Gamma}^{\rm s (c)} \hat{\chi}^{0} ( \bm{q} ) }
        .
\end{eqnarray}
 The ordered state is realized when the Stoner factor $\alpha_{\rm s (c)} = 1$, which is the maximum eigenvalue of $\hat{\Gamma}^{\rm s (c)} \hat{\chi}^{0} ( \bm{q} )$.
 Hereafter, we set the band filling as $n = 8.0$ and the temperature as $T = 0.02$~eV.
 We also set $J = U/6$ and $g = g'$.


\subsection{Spin fluctuations}
\begin{figure}[tb]
        \includegraphics[width=8cm]{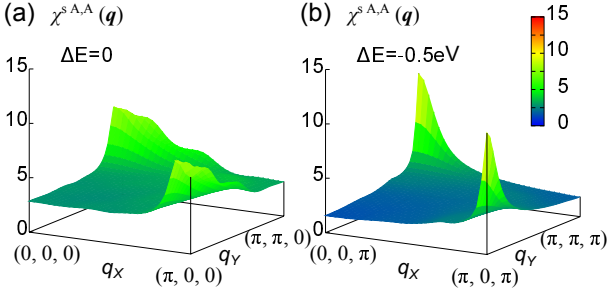}
        \caption{\label{fig4}
                (Color online)
                The obtained total spin susceptibility $\chi^{\rm s \, A A} (\bm{q})$ for (a) $\Delta E = 0$ and (b) $\Delta E = - 0.5$~eV.
        }
\end{figure}

 First, we calculate the total spin susceptibility resulting from Coulomb interaction:
\begin{eqnarray}
        \chi^{\rm s \, \alpha, \beta} ( \bm{q} )
        =
                \sum_{l,m}
                \chi^{\rm s \, \alpha, \beta}_{l, l, m, m} ( \bm{q} )
        .
\end{eqnarray}
 Figure~\ref{fig4}(a) shows the obtained $\chi^{\rm s \, A, A} ( \bm{q} )$ for $\Delta E = 0$ and $U = 2.32$~eV; the corresponding spin Stoner factor is $\alpha_{\rm s} = 0.98$.
$\chi^{\rm s \, A, A} ( \bm{q} )$ has its highest peak at the incommensurate wave vector $\bm{q} \approx ( \pi, 0, 0)$.
 Figure~\ref{fig4}(b) shows $\chi^{\rm s \, A, A} ( \bm{q} )$ for $\Delta E = - 0.5$~eV and $U_{\rm c} = 2.07$~eV ($\alpha_{\rm s} = 0.98$).
 In both cases, $\chi^{\rm s \, A, A} ( \bm{q} )$ has a peak at $\bm{q} = (\pi, 0)$.
 It seems that the intra-orbital nestings of Fermi surfaces,
  which give the enhanced spin susceptibility, are not significantly different in the two cases.
  (In contrast, Fe-based compounds has spin fluctuations with $\bm{q} = (\pi, \pi)$, \cite{Kuroki, Kontani_C66} which corresponds to the stripe-type SDW order. \cite{Rotter, Krellner, Huang})
 The obtained critical value $U_{\rm c} \ge 2$~eV is about twice as large as the value reported for iron-base superconductors using RPA calculations. \cite{Kuroki, Saito, Kontani_C66}
 This fact implies that a large Coulomb interaction is required to drive the SDW transition in BaNi$_2$As$_2$.
 We stress that the obtained $U_{\rm c}$ is largely underestimated since the self-energy correction is neglected.
 Therefore, we conclude that the spin fluctuations are very small in BaNi$_2$As$_2$.
 
\subsection{Orbital fluctuations}
\begin{figure}[tb]
        \includegraphics[width=8cm]{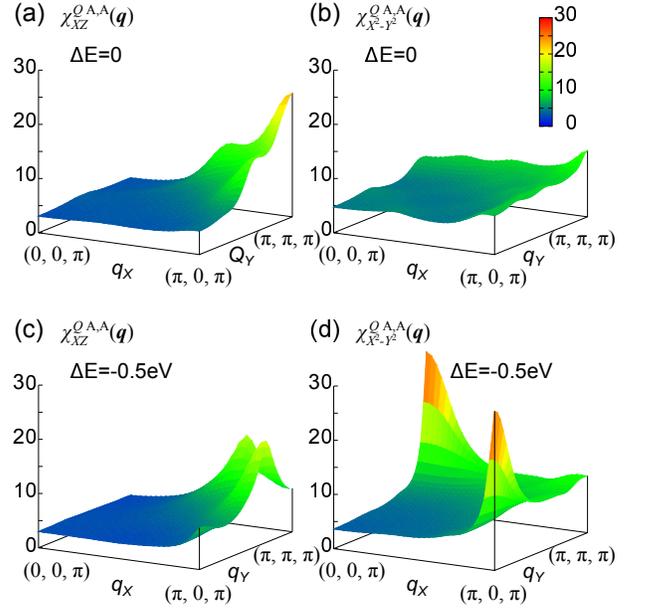}
        \caption{\label{fig5}
                (Color online)
                The charge quadrupole susceptibility $\chi^{\rm Q \, A A}_{\Gamma} (\bm{q}) $ for [(a) and (b)] $\Delta E = 0$ and [(c) and (d)] $\Delta E = - 0.5$~eV.
                The peak structure of $\chi_{X^2-Y^2} (\bm{q})$ for $\Delta E = - 0.5$~eV in (d) is consistent with the experimental zigzag structure in Fig.~\ref{fig1}.
        }
\end{figure}

 Next, we calculate the charge quadrupole susceptibilities resulting from the quadrupole interaction: \cite{Kontani_C66}
\begin{eqnarray}
        \chi^{\rm Q \, \alpha, \beta}_{\Gamma} ( \bm{q} )
        =
                \sum_{l, l', m, m'}
                o_{\Gamma}^{l, l'}
                \chi^{\rm c \, \alpha, \beta}_{l, l', m, m'} ( \bm{q} )
                o_{\Gamma}^{m, m'}.
\end{eqnarray}
 Figure~\ref{fig5}(a) and \ref{fig5}(b) show the obtained $\chi^{\rm Q \, A, A}_{XZ} ( \bm{q} )$ and $\chi^{\rm Q \, A, A}_{X^2-Y^2} ( \bm{q} )$ for $\Delta E = 0$, $U = 0$ and $g = 0.392$~eV, respectively.
 The corresponding charge Stoner factor is $\alpha_{\rm c} = 0.98$.
 When $\Delta E = 0$, $\chi^{\rm Q \, A, A}_{XZ (YZ)} ( \bm{q} )$ has its highest peak at $\bm{q} = (\pi, \pi, \pi)$,
  which is given by the inter-orbital nesting between the $d_{XZ (YZ)}$ orbital on FS$_2$ and the $d_{X^2-Y^2}$ orbital on FS$_3$.
 In addition, the inter-pocket nesting between different FS$_{\rm h}$'s also contributes to the $\chi^{\rm Q \, A, A}_{XZ (YZ)} (\pi, \pi, \pi)$. 
 The former and latter nesting vectors are shown as the thick and thin arrows in Fig.~\ref{fig3}(b), respectively. 
 The obtained peak position at $\bm{q} = (\pi, \pi, \pi)$, however, is inconsistent with the structural transition of BaNi$_2$(As$_{1-x}$P$_x$)$_2$.

 We also show $\chi^{\rm Q \, A, A}_{XZ} ( \bm{q} )$ and $\chi^{\rm Q \, A, A}_{X^2-Y^2} ( \bm{q} )$ for $\Delta E= - 0.5$~eV, $U = 0$ and $g = 0.399$~eV in Fig.~\ref{fig5}(c) and \ref{fig5}(d), respectively.
 The charge Stoner factor is $\alpha_{\rm c} = 0.98$.
 Compared to the case of $\Delta E = 0$, the peak of $\chi^{\rm Q \, A, A}_{XZ (YZ)} ( \bm{q} )$ at $\bm{q} = (\pi, \pi, \pi)$ is suppressed, and instead, $\chi^{\rm Q \, A, A}_{X^2-Y^2} ( \bm{q} )$ has its highest peak at $\bm{q} = (\pi, 0, \pi)$ (and $\bm{q} = (0, \pi, \pi)$) owing to the inter-orbital nesting between the $d_{X^2-Y^2}$ orbital on FS$_1$ and the $d_{3Z^2-R^2}$ orbital on FS$_{2}$.
 The nesting vector is shown as the thick arrow in Fig.~\ref{fig3}(c).
 The corresponding AF $O_{X^2-Y^2}$ order with $\bm{q} = (\pi, 0, \pi)$ is consistent with the zigzag chain structure reported experimentally, \cite{Sefat} as already shown in Fig.~\ref{fig1}.
 The obtained change in the inter-orbital nesting conditions originates from the noticeable shrinkage of the FS$_{\rm h}$. 

 The enhancement of $\chi^{\rm Q \, \alpha, \beta}_{\Gamma} ( \bm{q} )$ is mainly caused by the quadrupole interaction with respect to $o_{\rm \Gamma}$ in Eq.~(\ref{eq6}).
 Thus, $\chi^{\rm Q \, A, A}_{XZ (YZ)} ( \bm{q} )$ and $\chi^{\rm Q \, A, A}_{X^2-Y^2} ( \bm{q} )$ are mediated by the in- and out-of-plane oscillations of Ni ions, respectively. \cite{Kontani_SC, Saito, Onari_FLEX, Kontani_C66}
 In more detail,  both susceptibilities $\chi^{\rm Q \, A, A}_{XZ (YZ)} ( \bm{q} )$ and $\chi^{\rm Q \, A, A}_{X^2-Y^2} ( \bm{q} )$ enhance each other at $\bm{q} = (\pi, 0, \pi)$ through quadrupole off-diagonal term resulting from the orbital hybridization.

\section{Discussions}
 We have shown that experimental zigzag chain structure is reproduced by the divergence of $\chi^{\rm Q \, A, A}_{X^2-Y^2} (\pi, 0, \pi)$ due to quadrupole interaction $g$. 
 The obtained critical values $g_{\rm c} \sim 0.4$~eV for the quadrupole order are about twice as large as the values reported for Fe-based superconductors ($g_{\rm c}^{\rm Fe} \sim 0.2$~eV) using RPA calculations. \cite{Saito, Kontani_C66}
 However, a strong electron-phonon interaction of $\lambda_{\rm ep} = 0.76$ for BaNi$_2$As$_2$ was found using a DFT calculation, \cite{Subedi} whereas $\lambda_{\rm ep}^{\rm Fe} \sim 0.2$ for LaFeAsO. \cite{Boeri}
 Therefore, a charge quadrupole interaction mediated by an electron-phonon interaction would be dominant for BaNi$_2$(As$_{1-x}$P$_x$)$_2$, whereas the Coulomb interaction would be less important.
The origin of the zigzag chain structure 
is the AF $O_{X^2-Y^2}$ order induced by the 
inter-orbital nesting of the Fermi surfaces and quadrupole interaction.
Consistently, the total energy in the DFT is lowered 
by assuming the zigzag chain structure. \cite{Chen}

 Finally, we discuss the role of the AFQ fluctuations on the superconductivity.
 We have obtained  two types of AFQ fluctuations, $O_{XZ (YZ)}$ with $\bm{q} = (\pi, \pi, \pi)$ for $\Delta E = 0$ and $O_{X^2-Y^2}$ with $\bm{q} = (\pi, 0, \pi)$ for $\Delta E = - 0.5$~eV, in the BaNi$_2$As$_2$ system.
 Since $\Delta E = 0$ corresponds to the tetragonal phase, the AF $O_{XZ (YZ)}$ fluctuations would be dominant in the tetragonal phase.
 Thus, the superconductivity in the tetragonal phase with $T_{\rm c} \sim 3$~K recently reported by Kudo {\it et al.} \cite{Kudo} might be caused by the AF $O_{XZ (YZ)}$ fluctuations with $\bm{q} = (\pi, \pi, \pi)$. 
 In contrast, a large energy level shift $\Delta E$ would be required to realize the AF $O_{X^2-Y^2}$ fluctuations in the tetragonal phase.
 If we could realize the second-order structure transition in some way such as by applying uniaxial pressure, 
  the AF $O_{X^2-Y^2}$ fluctuations may be stronger,
  and higher-temperature superconductivity might be realized in Ni-based superconductors.

\section{Summary and future problems}
 To summarize, we investigated the electronic state and structure transition of BaNi$_2$As$_2$.
 The Coulomb and quadrupole-quadrupole interactions were treated within the RPA method. 
 The former interaction is not important since $U_{\rm c}$ is very large and, as a result, spin fluctuations are very small. 
 Owing to the latter interaction, two types of quadrupole fluctuations develop in BaNi$_2$(As$_{1-x}$P$_x$)$_2$: AF $O_{XZ (YZ)}$ and AF $O_{X^2-Y^2}$ fluctuations.
 The latter fluctuations are the origin of the experimental zigzag chain formation, \cite{Sefat}
  and the former fluctuations would then be the origin of the strong-coupling superconductivity observed in BaNi$_2$(As$_{1-x}$P$_x$)$_2$. \cite{Kudo} 
 $T_{\rm c}$ in Ni-based superconductors would increase further by developing  stronger AFQ fluctuations. 
 We conclude that both Ni- and Fe-based superconductors are characterized as strong orbital fluctuating metals, 
  although origins of their orbital fluctuations are different. 

In this study, we discussed the effect of the quadrupole interaction mediated by the electron-phonon interaction. 
However, we ignored the effects of the momentum dependence and differences between inter- and intra-plane oscillations,
 which would be important when several fluctuations develop simultaneously. 
Furthermore, we described the change of the band structure by introducing the parameter $\Delta E$ for simplicity. 
For more quantitative discussion, a high-accuracy tight-binding model is required.

 In the present paper, we discussed the zigzag chain formation due to the AF $O_{X^2-Y^2}$ fluctuations for $\Delta E = - 0.5$~eV, 
  and the resultant crystal structure is orthorhombic. 
 Experimentally, both longer bond ($\sim 3.1$~\AA{}) and shorter bond ($\sim 2.8$~\AA{}) are slightly modulated in a staggered way. \cite{Sefat}
  and the crystal structure becomes triclinic. 
 The origin of this triclinic structure formation is an important future problem. 
 
 We also comment on the role of VC: 
 In Ref.~\onlinecite{Onari_VC}, we found that the Aslamazov-Larkin-type VC due to Coulomb interaction causes strong orbital fluctuations, 
  which would be the main origin of the orthorhombic phase transition as well as superconductivity in Fe-based superconductors. 
 However, this mechanism will be unimportant in BaNi$_2$(As$_{1-x}$P$_x$)$_2$ because of the large $U_{\rm c}$. 
 On the other hand, the VC due to electron-phonon interaction might be important in BaNi$_2$(As$_{1-x}$P$_x$)$_2$. 
 By this mechanism, the AF $O_{\Gamma}$ fluctuations give rise to the ferro $O_{\Gamma'}$ order if ${\rm Tr} \{ O_{\Gamma}^2 O_{\Gamma'} \} \ne 0$ (two-orbiton term). \cite{Onari_VC, Kontani_C66}
 In the present model, this mechanism can realize the ferro $O_{3Z^2-R^2}$ order for $\Delta E = - 0.5$~eV, 
  which causes the change in the inter-layer distance with preserving the tetragonal symmetry. 
 It might be related to the abrupt change in $c$-axis resistivity at $T_{\rm s}$. \cite{Sefat} 
 This is another important future problem.

\acknowledgments
We thank T. Saito for the useful discussion on the RPA calculations and the program development.
Numerical computation in this work was partially carried out at the Yukawa Institute Computer Facility.
This study has been supported by Grants-in-Aid for Scientific Research from MEXT, Japan, and by JST, TRIP.
This work was also supported by a Grant-in-Aid for Scientific Research on Innovative Areas ``Heavy Electrons" (No. 20102008) of the Ministry of Education, Culture, Sports, Science, and Technology, Japan.


\end{document}